\documentclass{PoS}

\title{Solar panels as cosmic-ray detectors}

\ShortTitle{Solar panels as cosmic-ray detectors}

\author{\speaker{Carlo Stella}\,\thanks{E-mail: {karlo.stella@gmail.com}}\,, Michele Palatiello\\
Universit\`a di Udine and INFN  Trieste, Gruppo Collegato di Udine, Italy}

        \author{Pedro Assis, Pedro Brogueira, Catarina Esp\'irito Santo, Patricia Gon\c{c}alves, M\'ario Pimenta\\
        LIP - Laborat\'{o}rio de Instrumenta\c{c}\~{a}o e F\'{i}sica Experimental de Part\'{i}culas, Lisboa, Portugal}

\author{Alessandro De Angelis\\
        INFN Padova, Italy, and LIP/IST, Lisboa, Portugal}

\abstract{Due to fundamental limitations of accelerators, only cosmic rays can give access to centre-of-mass energies more than one order of magnitude above those reached at the LHC. In fact, extreme energy cosmic rays ($10^{18}$ eV - $10^{20}$ eV) are the only possibility to explore the 100 TeV energy scale in the years to come. This leap by one order of magnitude gives a unique way to open new horizons: new families of particles, new physics scales, in-depth investigations of the Lorentz symmetries. However, the flux of cosmic rays decreases rapidly, being less than one particle per square kilometer per year above $10^{19}$ eV: one needs to sample large surfaces. A way to develop large-effective area, low cost, detectors, is to build a solar panel-based device which can be used in parallel for power generation and Cherenkov light detection. Using solar panels for Cherenkov light detection would combine power generation and a non-standard detection device.}

\FullConference{Science with the New Generation of High Energy Gamma-ray experiments, 10th Workshop \\
                 04-06 June 2014\\
                 Lisbon - Portugal}

\begin{document}

\section{Introduction}

In the first half of the $20^{th}$ century, the history of particle physics could hardly be distinguished from the history of cosmic rays; with the advent of accelerators, however, cosmic rays were almost ``forgotten'' by the particle physics community. The Large Hadron Collider (LHC) at CERN, a superb realization of physics and engineering, culminates a brilliant path of more than fifty years in which an increase of at least one order of magnitude in the centre-of-mass energy of the accelerators was achieved every ten years. New physics scales were systematically explored; the beautiful Standard Model of particle physics was discovered and impressively tested. 

The field of extreme energy cosmic rays was born half a century ago, with the detection of the first air shower corresponding to a primary particle energy of about $10^{20}$ eV by John Linsley, at the Volcano Ranch experiment \cite{Volcano}, in 1962. In the following decades, several experiments contributed to slowly consolidate the evidences for the existence of such high-energy particles, and to develop detection and reconstruction techniques. While the origin and acceleration mechanisms of such high-energy cosmic particles remain a challenging astrophysical question, the nature of these particles and their interactions in the atmosphere address fundamental particle physics questions. The centre-of-mass energies involved in these interactions are more than one order of magnitude above those reached at the LHC. Extreme energy cosmic rays are the only possibility to explore the 100 TeV energy scale in the years to come. This leap by one order of magnitude gives the only possibility to open new horizons: new families of particles, new physics scales, in-depth investigations of the Lorentz symmetries. 

This article aims to indicate a possible route to perform such an exploration. It follows unconventional approaches, choosing techniques that are non-standard in air-shower detection and that are of interest for other domains of application. It is intrinsically interdisciplinary, as it evolves in the borderline between particle physics, astrophysics and cosmic ray physics.

\section{Detection of cosmic particles at extremely high energies}

The scarce flux of cosmic rays at extreme energies (about one per km$^2$ per year above $10^{19}$ eV) makes the direct detection of these particles in balloons or satellites impossible. Extreme energy cosmic rays must be studied through the extensive air showers originated by their interaction with the atmosphere, and the subsequent multiplication of particles \cite{HECR}. Different types of primary particles generate air showers with different characteristics. Most of the air shower content at ground is made of electrons, photons, and muons, creating a pool of particles of several kilometers in diameter. 

Showers can be detected by sampling the different kind of lower energy secondary particles reaching the ground, and high-energy air shower experiments must cover huge detection areas to compensate the limited flux of extreme energy cosmic rays. For this reason the choice of detectors is a trade off between cost and performance, namely sensitivity and precision. To detect directly charged particles in the showers, usual choices are scintillators and Cherenkov water tanks. In order to cover large areas, a relatively simple detection unit is replicated hundreds or thousands of times. For practical reasons, such as cost and maintenance logistics, these detector units are usually simple and the level of detail with which showers are measured is limited.

Showers can also be studied through the detection of their electromagnetic radiation component emitted at different frequency and with different mechanisms:

\begin{itemize}
  \item Fluorescence detectors measure the longitudinal profile of the shower, from its growing stages, up in the atmosphere, to its fading out close to ground. Images of the shower track are formed from long distances, by focusing the ultraviolet light (UV) emitted by nitrogen molecules excited by the passage of charged particles belonging to the cascade. Since the light emission is isotropic, and proportional to the energy deposited, this technique gives a very good measurement of the total number of charged particles as the shower evolves. This detection is limited by the efficiency of UV light production and by the sensitivity and coverage of current detectors.
  \item Ultra-relativistic shower particles travelling in the atmosphere also emit Cherenkov radiation, collimated along the shower axis. Shower Cherenkov light detection is the basis of TeV gamma ray telescopes.
  \item Showers also emit  radiation in the MHz range, that can be recorded by arrays of antennas at ground. This technique is still in development and, with the current prospects for obtaining measurements of the shower arrival direction and energy, might play an important role on the study of the astrophysical source of extreme energy cosmic rays. 
\end{itemize}

\section{Solar panels as cosmic ray detectors}

In a cosmic ray detector array, each station has to be powered, and it is common that it is equipped with a set of photovoltaic solar panels. Since solar cells are n-i-p junctions with high quantum efficiency, covering the spectral range of Cherenkov light, they may also be used to detect the Cherenkov photon pulse emitted in the atmosphere by showers above $10^{18}$ eV \cite{Solar1} (wavelength from 300nm to 1000nm, high intensity and a narrow time spread of about 10ns). Indeed the possibility of using solar panels for this purpose was explored in the end of the '90s \cite{Solar2}. The development of technology and the particular case study of this article make the choice particularly interesting today.

\subsection{Methodology}

We assume as typical conservative scenario the detection of showers at a distance of 500 m from the core (the distance between ground stations is usually 1000-1500 m), an efficiency of 15\% for the solar cell, which can be reached by commercial photovoltaic panels \cite{Solar3}, an integration time of 50-100 ns for the signal and neglect the noise introduced by the electronics. In a proton-initiated shower of $10^{18}$ eV (1EeV) one expects from 1 to 2 Mphotons/m$^2$ (depending on the zenith angle, from 0 to 60 degrees) from Cherenkov radiation in the visible at a distance of 500 m from the shower core. This yield grows linearly with energy. The FWHM duration of the signal is of the order of 10-100 ns, depending on the zenith angle. Full moon yields about 1000 Mphotons/m$^2$ in 100 ns in the visible. The intrinsic energy threshold for night time observation ($S/\sqrt B = 3$ \cite{NSB}) at 500 m from the core can thus, taking into account the efficiency $\epsilon$ of the detector, be parametrized as:

\begin{equation}
E_{thr} (EeV) \sim 0.08 \cdot I \cdot \sqrt{ (5 m^2 / A) (30\% / \epsilon) }
\end{equation}
where $I$ is the square root of the number of background photons per square meter in the visible, normalized to the full moon condition ($I=1$). This translates into a threshold of about 0.11 EeV with full moon; about 0.04 EeV in dark nights; about 0.4 EeV in dusk/dawn conditions; about 10 EeV with average sunlight. Reasonably one can thus assume a threshold of some $3.10^{17}$ eV with a 50\% duty cycle. For energies above $10^{19}$ eV, duty cycles of 100\% may be reachable (also taking into account that the signal at 300 m from the shower core is 5 times larger than at 500 m).

In addition, if a time resolution O(20 ns) can be reached, the standard deviation in the arrival time for photons from a particular height can be translated back into a relationship between time and height in the mid-atmosphere, with an uncertainty of about $\pm$35 g/cm$^2$ only \cite{Solar1}. Finally, one can hope for an increase of sensitivity due to the development of multi-junction solar cells in the next few years.

The use of large solar panels as Cherenkov light detector favour the signal to noise ratio (as it scales with the square root of the collection area). However, detectors with large areas result in large capacitances and require special care in the measuring design, namely in the coupling match between the solar panel and the test equipment that optimize filtering the faint light fast signals from the background. Laboratory tests of solar cells reported in the year 2000 \cite{Bologna} suggest that a signal from Cherenkov light can be extracted from solar panels even during daylight by the use of inductive coupling.

We propose to further develop and test this concept profiting from the recent developments in solar panels \cite{Solar4}. We will assess and test in laboratory several panels available in the market based in different technologies such as silicon, polycrystalline thin films and multi-junction to find the best compromise between price, power generation and Cherenkov light detection. For instance, multi-junction cells \cite{Solar5} may present a good alternative as, although expensive, they present better efficiency for power generation (reaching values around 40\% for the Concentrating Triple Junction design) and are sensitive to a spectrum extended to the UV, allowing harnessing this part of the Cherenkov light, increasing the output signal. In the laboratory, panels will be tested using calibrated LEDs to generate light pulses of several wavelengths with $10^7 \sim 10^{10}$ photons spread over a time of tens to hundreds of ns to emulate Cherenkov light while a second light source will be used to emulate the background light. Later the panels will be tested under field conditions and in coincidence with a particle ground array detector to allow for cross-calibration. 

\subsection{Detection unit design}

\begin{figure}[htbp]
  \begin{center}
  \includegraphics[width=14cm]{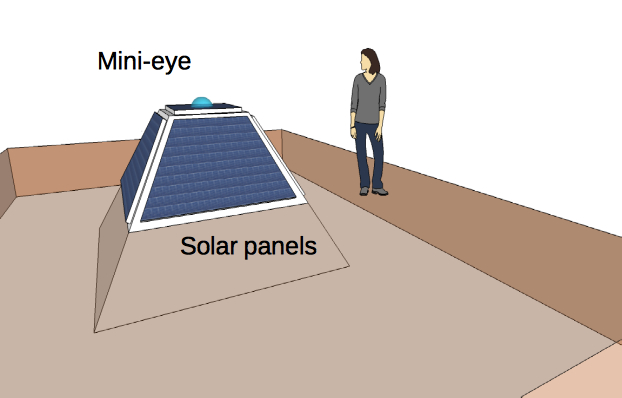}
  \caption{Solar panel unit prototype\label{fig:spup}}
  \end{center}
\end{figure}

The solar panel-based detector set-up for the detection of Cherenkov light from EAS will be placed on a truncated pyramid setup (Fig.\ref{fig:spup}), since this improves the geometrical efficiency. The solar panel assembly consists of 5 panels, one placed horizontally on the top of the structure and the remaining four panels covering the structure from the sides, down to ground level. The baseline configuration uded for the simulation consists of 9 m$^2$ of solar panels. The time resolution of current technology solar panels is above 5000 ns, setting a limit on the reconstruction of the shower variables from Cherenkov photons using solar panels only. As discussed above, the Cherenkov pulse signal can be detected under night sky background conditions, including full moon, but also under dusk background conditions. The duty cycle of the panel set up is expected to be greater than 50\%. Additionally, the solar panel supplies the necessary power for the complete integrated system. 

A SiPM-based mini-eye with full sky coverage might be placed on top of truncated pyramid. The advantages of using the SiPM mini-eye in addition to the solar panel are its fast response time, O(10ns), and the  angular resolution that can be achieved with the current set-up. The SiPM mini eye is placed on top of the central solar panel. The SiPM mini-eye is designed to detect Cherenkov signals from EAS under Night Sky Background conditions. Under the same conditions, it will also be sensitive to the shower fluorescence light. Given their characteristics, the optimization of the solar panel geometric configuration and of the required SiPM mini eye parameters shall be performed simultaneously. 

\section{Outlook}
The goal of designing and building an integrated detector unit including all the developed technologies for a ``real life'' test-bench scenario, it is a challenge related to the integration of the different parts: geometry, electronics, shadow/shielding effects, design optimization, mounting logistics. But from the point of view of cosmic-ray studies, the global optimization of the design parameters and the performance studies can be a major contribution for the next generation of cosmic ray detectors.

\end{document}